When Costly Punishment Becomes Evolutionarily Beneficial

By

Ivan C. Ezeigbo
Institution: Minerva Schools at KGI

## Abstract

Cooperation in evolutionary biology means paying a cost, c, to enjoy benefits, b. A defector is one who does not pay any cost but enjoys the benefits of cooperators. Human societies, especially, have evolved a strategy to discourage defection – punishment. Costly punishment is a type of punishment where an agent in a biological network or some cooperative scheme pays a cost to ensure another agent incurs some cost. A previous study shows how parameters like diversity in neighbors across games, density of connectivity, and costly punishment influence the evolution of cooperation in non-regular networks. In this study, evolution in regular networks due to the influence of costly punishment is also considered, specifically spatial lattices. This study compares observations between non-regular and regular networks as these parameters change and brings a clearer understanding of interactions that occur in these networks. The models, results and analysis bring a new understanding to game theory and punishment. The results show that costly punishment never arises as a sole evolutionary strategy. However, in evolutionary games where costly punishers could evolve more favorable strategies, the initial presence of costly punishers would bring about high average payoffs in all types of regular networks and heterogeneous networks. In regular networks, every node has the same degree, $k$. Although punishment is conventionally thought to be anti-progressive, in the presence of diversity in neighbors, it magnifies the payoff of a group for all heterogeneous networks. In regular networks however, diversity in neighbors is not required for costly punishment to boost average payoff. This suggests an answer to the question on why costly punishment has been favored by natural selection, which is particularly obvious in human populations.

**Significance Statement:** In this paper, I report on how different network structures differentially influence evolutionary behaviors like cooperation. This paper discusses interesting interactions that occur in two types of networks - regular and non-regular networks - but most importantly, it poses an interesting theory from the experimental results gathered. It is still a conundrum as to why people would incur a cost to punish cheaters or defectors. I show that although punishment is conventionally thought to be anti-progressive, in the presence of diversity, it magnifies the payoff of a group. This poses a possible answer to the puzzle on why costly punishment has been favored by natural selection, which may as well explain our impressive evolutionary advancement and exceptional ability to cooperate like no other species on this planet. Indeed, diversity and punishment in human societies may actually explain our evolutionary advancement.

## Introduction

Costly punishment, the strategy of paying a cost to decrease the payoff of another agent, has proven to be effective against defection in many studies (1-6). A cooperator would pay a cost for some benefit. Defectors expect benefit in a cooperative scheme, without paying any cost. Costly punishers on the other hand would pay some cost to reduce the payoffs of another agent or strategy, by purpose, usually the defective strategy (7). Costly punishment was shown to disfavor defection in networks in favor of an equilibrium state of cooperators, defectors and punishers, but did not necessarily favor cooperation, which resulted in a decrease in average payoff. However, when diversity in neighbors was introduced into the network, both defection and the emergence of cooperation were favored. Whereas the defective strategy was only bolstered, cooperation emerged as a similarly competitive sole strategy (5). This is in agreement with other studies that have demonstrated that punishment could favor cooperation (1-4, 8-10) because, essentially, the disfavoring of defection in the local level of analysis is equivalent to the favoring of cooperation (5). With reference to a previous study (5), we see in this study that spatial lattices exhibit some similar pattern to the influence of costly punishment; defection is slightly bolstered, and cooperation emerges in the presence of diversity in neighbors. In this study, an agent-based model is designed to compare the influence of regular structural networks, like spatial lattices, on the evolution of gaming strategies to interactions seen in non-regular networks. There are certain interesting relationships that are gleaned from this study: 1) Spatial square lattices produce the same results as costly punishment but annuls the effect of costly punishment on defective strategies. Spatial square lattices disfavor defection, as do costly punishment; however, when a non-regular network structure with punishers is made regular – in this case, made into a spatial square lattice – defection is favored, and average payoff rises. There are several research that suggest a possible influence of structured regular networks on defection (11-14). 2) Punishment never arises as the sole evolutionary strategy, as shown from the experimental data. Costly punishment, whether in a regular structure or non-regular structure, never emerges as the sole evolutionary strategy. This is sensible as punishment is a strategy that may have evolved for the purposes of discouraging defection. It is not a sustaining strategy. This is in agreement with the findings of Dreber et al (15). Although costly punishment may exist in equilibrium with other strategies (that must include defectors) in non-regular networks, it never arises as the sole evolutionary strategy. Costly punishment would only coexist in equilibrium only in the presence of defectors. Dreber et al (15) have shown in their study that costly punishment leads to low average payoffs. Different experimental designs are used to validate this claim for non-regular network structures. However, this is not valid for spatial lattices with punishers in evolutionary games (where a punisher can evolve a more profiting strategy at any time in the game). The zero probability of the sole emergence of punishment in an evolutionary game is a consequence of the possibility of punishers evolving better strategies after they are introduced into the network. In regular structures however, introducing costly punishers bolsters average payoff of the network. 3) After a game is run in the absence of diversity in neighbors and an equilibrium state is attained, costly punishers are always absent in this equilibrium state in spatial square lattices but may be present in non-regular structures. In other words, the evolution of an equilibrium state guarantees the absence of costly punishers in square lattices, but the same does not apply to non-regular structures. This further strengthens the observation on how regular structures like square lattices are not only anti-defective but also anti-punishing. In other words, these regular structures strip away non-sustaining strategies (11-14). 4) A previous study showed that introducing diversity in neighbors to any network converts its equilibrium state to a cooperative strategy (5). In this study, we see that introducing diversity in neighbors to non-regular network with no costly punishers does not only convert its equilibrium state to a cooperative strategy but also disfavors defection. This reduced fraction of the defective strategy is as well converted to a cooperative strategy. There are several studies that support the evolution of cooperation in biological networks following the introduction of diversity to those networks (16-18). In the presence of costly punishment in non-regular network structures, we expect defection to be favored, rather than disfavored, with the introduction of diversity in neighbors, as a previous study shows how diversity in neighbors boosts both the defective strategy and the emergence of cooperation in the more global scale of the evolutionary game. In regular network structures like square lattices with costly punishment, defection is rather disfavored in favor of cooperation when diversity in neighbors is introduced, as is shown in this study.

**Experimental Design**

In this study, two types of networks are of primary focus here – a regular square lattice and a non-regular network. In the regular square lattice, each agent in the network is connected to 4 other agents in the network, and with the influence of diversity in neighbors, these neighbors could change but the number of neighbors, $k = 4$, remains the same for every agent. This is a spatial structure and each agent engages in a multiplayer Prisoner's Dilemma game with their four neighbors. The rewards and payoffs are calculated using the payoff matrix in Dreber et al. (15). When $N$ is the total number of agents in the network, an agent in a non-regular network could be connected to at most $N - 1$ agents and at least 1 agent to ensure the network remains a connected graph. Hence, the value of $k$ would vary for each player, unlike the regular spatial lattice as neighbors, as well as the neighbors themselves; these are assigned randomly. With the influence of diversity in neighbors, the number of neighbors, $k_i$, as well as the neighbors themselves could change at each round of a game. The density of connectivity, $D_c$, is the total number of edges or connections in the network. It is given by the formula,

$$D_c = N \times f, \quad where\ f\ is\ a\ multiplying\ factor.$$

In a regular square lattice, we have our density of connectivity to be

$$D_c = \frac{kN}{2} = \frac{4N}{2} = 2N, \quad where\ k = 4.$$

This makes sense as the sum total of edges for the square lattice of $k = 4$ since we have to avoid repetitions that arise from having two ends of an edge; that is, agent A connected to agent B is exactly the same as agent B connected to agent A. However, in a non-regular network we would have our density of connectivity to be

$$D_c = \frac{1}{2} \sum_{i=1}^{N} k_i\ , \quad where\ 1 \leq k_i \leq N - 1.$$

Experiments carried out in this study compares evolutionary strategies between both networks when they have the same density of connectivity, $D_c$, and the same number of agents, $N$. In other words, we consider cases when

$$D_c = Nf = 2N = \frac{1}{2} \sum_{i=1}^{N} k_i$$

From $N \times f = 2N$, we can see that for any value of $N$ we consider for the non-regular network, $f$ should remain constant at $f = 2$. The experiments were carried out using agent-based models that ran 100 games, with each game having 1000 rounds. Agents associate with other agents in Prisoner's Dilemma multiplayer games and payoffs are calculated from each interaction using the payoff matrix from Dreber et al. (15). Average payoff, $P_{avg}$, in this experiment, when both types of networks have the same density of connectivity, $D_c$, and the same number of agents, $N$, was hence calculated as follows:

$$P_{avg} = \frac{1}{N} \sum_{i=1}^{4N} P_i, \quad where\ P_i\ is\ the\ payoff\ of\ agent, A, from\ interaction, i.$$

Thus, the total payoff, $P_A$, of a particular agent, $A$, from its $k_A$ different interactions would be:

$$P_A = \sum_{i=1}^{k_A} P_i$$

Consequently,

$$P_{avg} = \frac{1}{N} \sum_{i=1}^{4N} P_i = \frac{1}{N} \sum_{A=1}^{N} P_A = \frac{1}{N} \sum_{A=1}^{N} \sum_{i=1}^{k_A} P_i$$

Obviously, for a regular spatial lattice, $k_A = 4$. This study shows how the rearrangement or reconstruction of a biological network with constant properties, $N$ and $D_c$, from non-regular network to a regular structural lattice can influence not only the degree of cooperation or defection observed but the average payoff of the network as well. Figure 1a – d shows the structure of two non-regular networks of N = 25 and N = 64, gotten from the model. Figure 1e – f however, shows the diagrammatic representation of a regular spatial lattice structure.

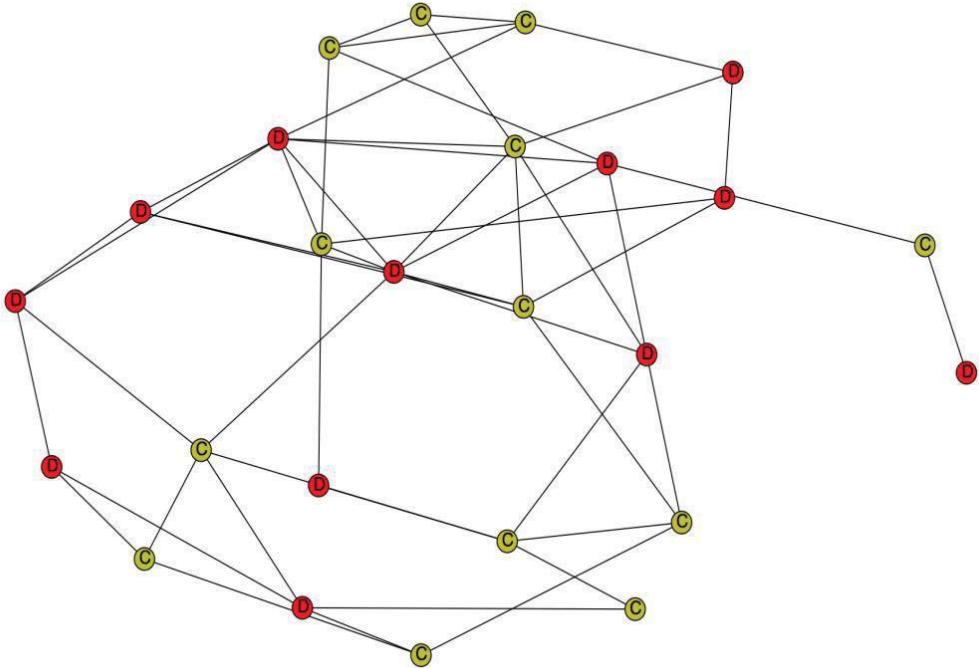

**Figure 1a:** Non-regular network of N = 25 and f = 2.0 without costly punishers. Cooperators are yellow nodes, defectors are red nodes.

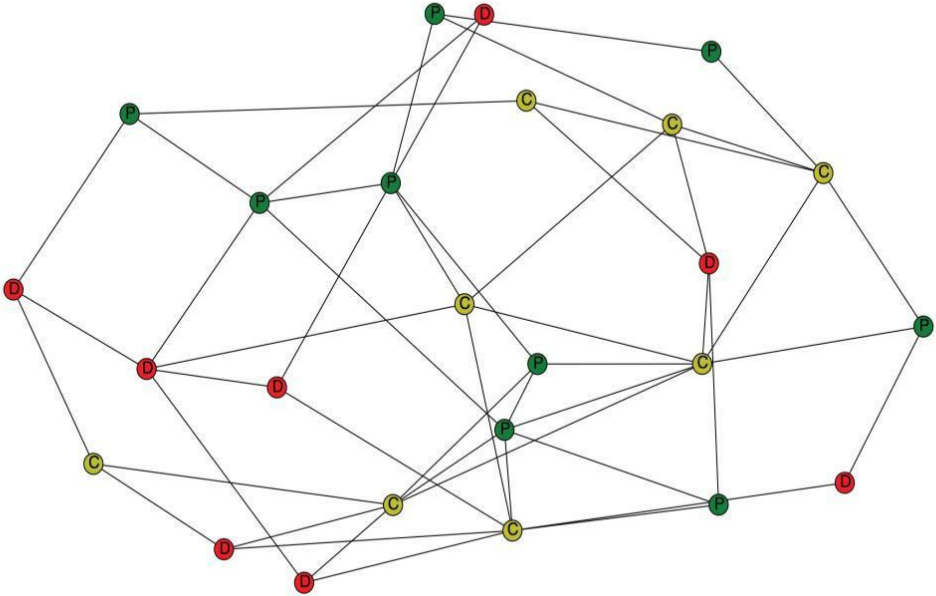

**Figure 1b:** Non-regular network of N = 25 and f = 2.0 with costly punishers. Costly punishers are green nodes.

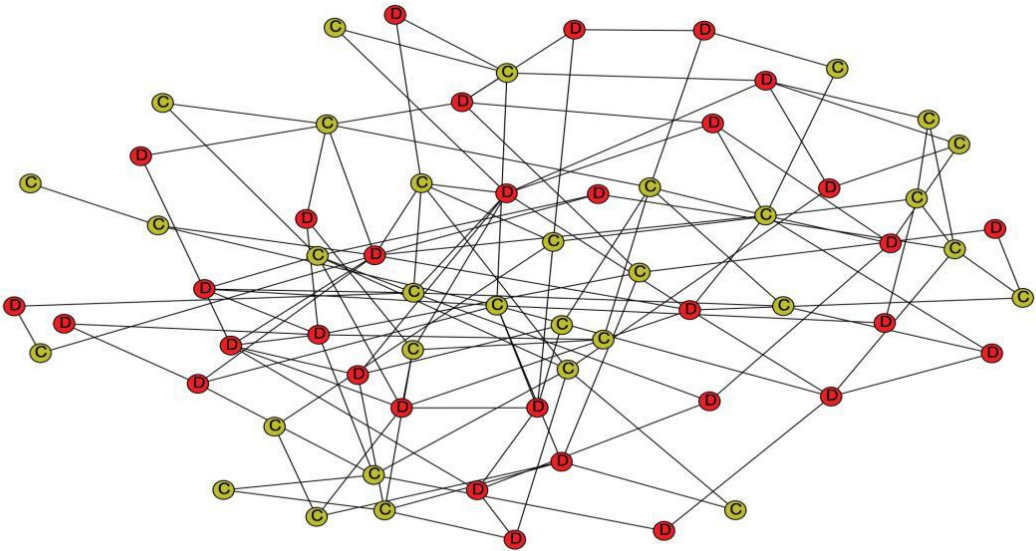

**Figure 1c:** Non-regular network of N = 64 and f = 2.0 without costly punishers.

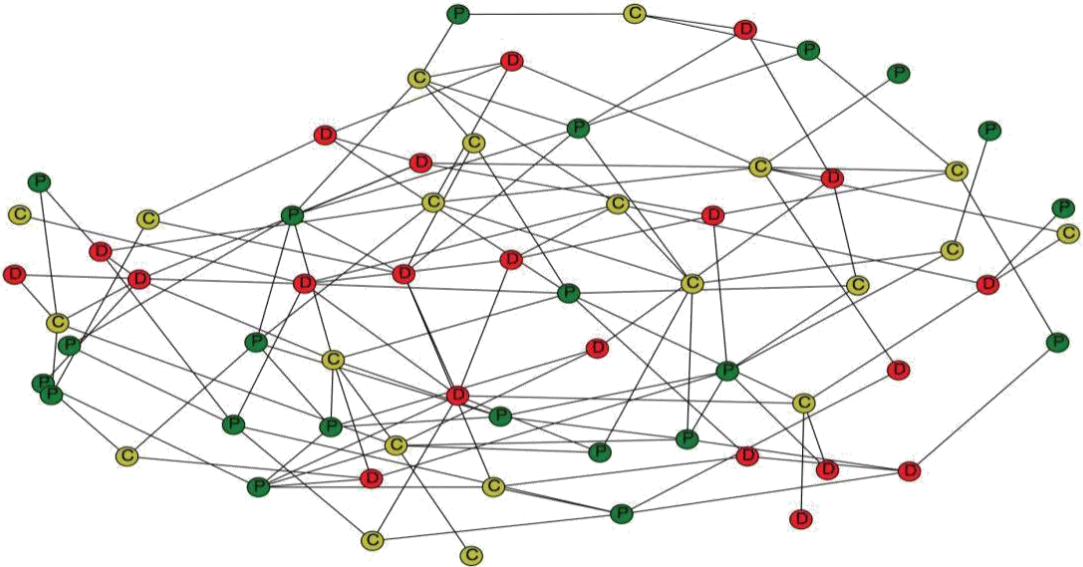

**Figure 1d:** Non-regular network of N = 64 and f = 2.0 with costly punishers.

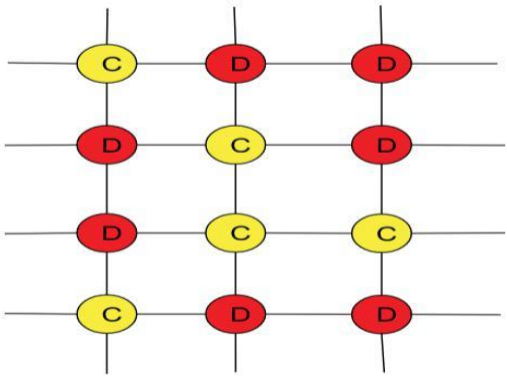

**Figure 1e:** Diagrammatic representation of a subnetwork of a spatial square lattice without costly punishers. Yellow nodes are cooperators while red nodes are defectors. Every agent in the network is connected to four other agents. This is what makes them regular.

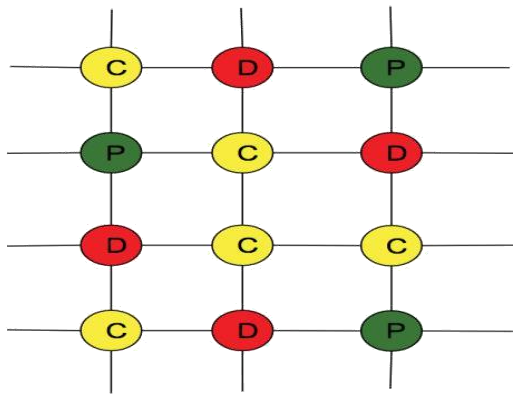

**Figure 1f:** Diagrammatic representation of a subnetwork of a spatial square lattice with costly punishers. Green nodes are costly punishers. Every agent in the network is connected to four other agents.

The following key results were observed.

## Results

The following relations in regular and non-regular networks are observed:

1. **Spatial lattices produce the same results as costly punishment but annuls the effect of costly punishment on defective strategies**

It is observed that, like costly punishment in non-regular networks, the presence of a regular structure like a spatial lattice is enough to disfavor defection, and most importantly, in the presence of diversity in neighbors, spatial lattices would favor both the emergence of cooperation and the increase in defection of the network. This influence of spatial structures on the evolution of cooperation is well explored in the literature (11-12, 14). In non-regular structures with costly punishers, I have shown in previous study that the introduction of diversity in neighbors would promote defection and the emergence of cooperation (5). These patterns illustrate how regular structures like spatial lattices show a similar effect to costly punishment in non-regular networks. Network payoffs also appear higher on average with regular structures as compared to non-regular structures, before the introduction of costly punishment and diversity in neighbors (see Table I, II, III, & IV). This is an interesting observation with regular structures as it brings to light that rearranging non-regular network to a regular structure can affect what strategies evolve and even influence average payoff. However, on rearranging a non-regular network with costly punishers to a regular structure, we do not observe an increase in the anti-defective characteristics of spatial lattices; rather we observe a counter effect on defection in this complex network. There is a rise in defection and average payoff; there is a slight decrease in average payoff and defection in the presence of diversity in neighbors, nonetheless, when compared to homogeneous regular networks. We only observe a significant amplification in the anti-defective characteristics of spatial lattices when costly punishment is introduced in a regular network; nevertheless, we still see a rise in average payoff both in the presence and absence of diversity in neighbors. As emphasized in a previous study, interpretation of the increase or decrease in strategies and payoffs is relative and would differ with the initial state of the network (5). In this case, if the initial state is a homogenous network with costly punishers, then the consequence of introducing a regular structure would result in an increase in the probability of defection; whereas, if the initial state was a homogenous network with a regular structure, then the consequence of the introduction of costly punishers would be a decrease in the probability of defection. This is a potential reason for much discrepancy in the literature on the effects of diversity in neighbors and regular structures on the evolution of cooperation in a biological network given that there is some complex network property that exists between the costly punishment strategy and the regular structure (19-24). Essentially, researchers interpreting their results from different frames of reference may render conclusions faulted by this oversight; the order of introduction of costly punishers and regularization may affect how one interprets the effect of regularizing non-regular network. Figure 2 and 3 shows the decrease in defection for two networks, N = 25 and N = 64, when the homogenous non-regular network is regularized. Figure 2 shows results for non-regular network while Figure 3 shows results for their respective regularized networks.

```
('Cope:', 0, ' Defe:', 100)
('Average payoff is', 31.35)
('Total agents', 25)
```

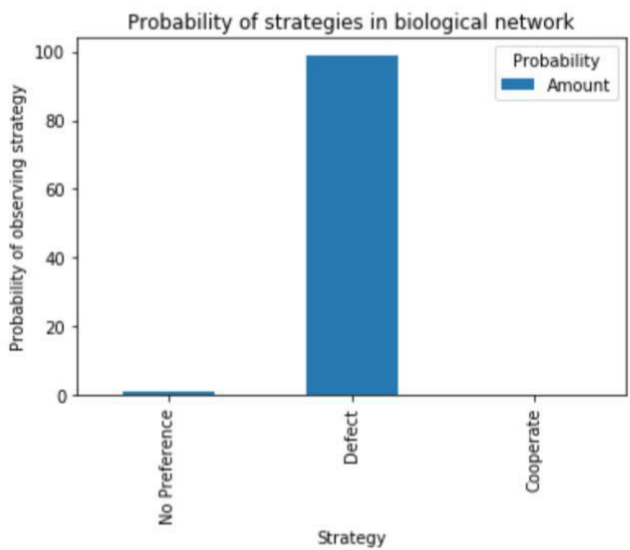

```
Counter({'Defect': 99, 'No Preference': 1, 'Cooperate': 0})
```

**Fig 2a:** The probability of defection in a homogenous, non-regular network of N = 25 is about 99%, while about 1% is the probability of achieving an equilibrium state. The probability of cooperation is 0%. 'Cope' and 'Defe' shows the fraction of times people cooperated in the major part of a game and the fraction of times people defected in the major part of the game respectively. Note that this may not always add up to 100 since people may have as well punished in the major part of the game.

```
('Cope:', 0, ' Defe:', 100)
('Average payoff is', 34.93)
('Total agents', 64)
```

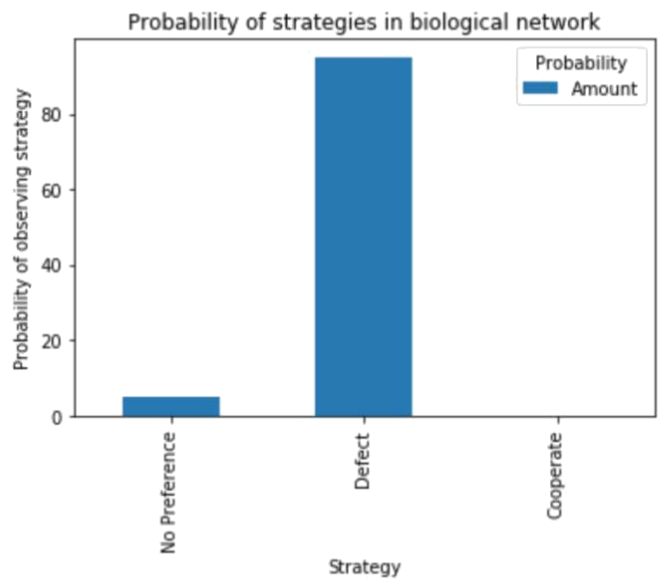

```
Counter({'Defect': 95, 'No Preference': 5, 'Cooperate': 0})
```

**Figure 2b:** The figure shows the result for N = 64. As shown, the probability of defection arising as the sole evolutionary strategy is 95% and there is a 5% chance that we achieve an equilibrium/ No Preference state.

```
('Cope:', 0, ' Defe:', 100)
('Average payoff is', 51.74)
('Total agents', 25)
```

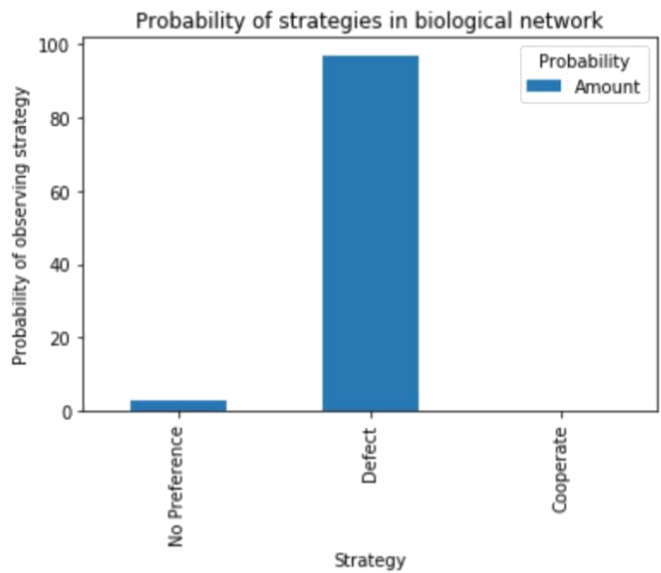

```
Counter({'Defect': 97, 'No Preference': 3, 'Cooperate': 0})
```

**Figure 3a:** The network in Fig. 2 is regularized to a spatial lattice, while keeping the density of connectivity constant, and we see a decrease in defection from 99% to 97%. There is a subsequent rise in the equilibrium state as well.

```
('Cope:', 0, ' Defe:', 100)
('Average payoff is', 116.4)
('Total agents', 64)
```

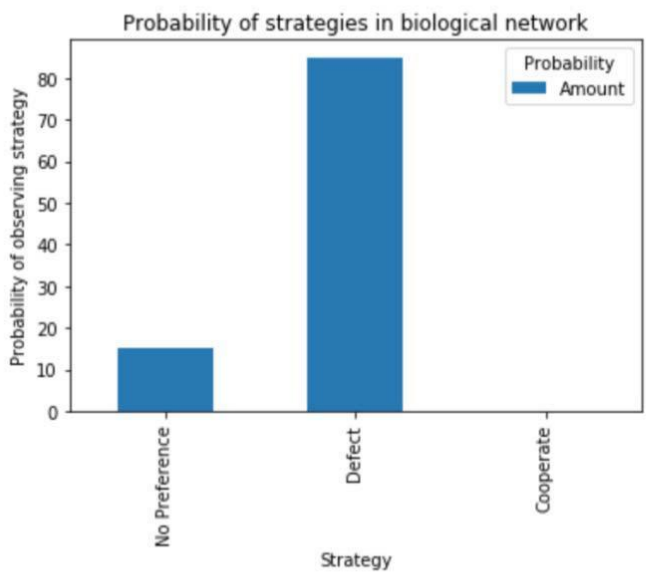

```
Counter({'Defect': 85, 'No Preference': 15, 'Cooperate': 0})
```

**Figure 3b:** The figure shows the result for the regularization of N = 64. There is a reduction in defection from 95% to 85% and a subsequent rise in the equilibrium state from 5% to 15%.

## 2. Punishment is never a globally evolved strategy

Winners don't punish. This was demonstrated by Dreber et al. (15) who were able to show that agents with the highest payoff do not participate in costly punishment – essentially, winners don't punish (5). This observation is further buttressed by results from the model that the costly punishment strategy, in both regular and non-regular networks, never arises as the sole evolutionary strategy. Here, it is important to draw a clear distinction between two categories of strategies based on their evolutionary consequences. These are *sustainable strategies* and *sustaining strategies*. Sustainable strategies are strategies an agent adopts for sustenance in the presence of one or more other biological agents. It is a strategy that can thrive on its own. Cooperation is a sustainable strategy. A network consisting entirely of cooperators would continually thrive unless the benefits of cooperation are less than the cost for cooperating. Ohtsuki et al. (7) showed that the cooperative strategy would thrive when $\frac{b}{c} > k$, such that b is the benefit, c is the cost and k is the average number of neighbors in the network. Defection and punishment are not sustainable strategies because a network consisting entirely of defectors or punishers cannot experience growth in payoffs. As a matter of fact, a network consisting entirely of costly punishers may retrograde rather than progress. A sustaining strategy on the other hand is a strategy an agent adopts to contribute to its payoff and sustain itself. A strategy may be sustaining but not sustainable. Defection is an example of such strategy. Defectors grow and thrive at the expense of cooperators but cannot grow and thrive by themselves. Hence, they sustain themselves at the cost of another agent. Costly punishment is neither a sustainable nor sustaining strategy. Costly punishers are not interested in personal gain; rather they are sometimes altruistic and exist only for the purposes of reducing the fitness of defectors in the network (1-2, 4-6). Cooperation is both a sustainable and sustaining strategy since it contributes a benefit in payoff to the agent. All sustainable strategies are expected to be sustaining strategies, and non-sustaining strategies are usually non-sustainable. This helps illuminate why costly punishment never emerges as the sole evolutionary strategy. It is clearly not a sustaining strategy. Only sustaining strategies can emerge as sole evolutionary strategies because the entire purpose of evolution is to favor the fittest. However, in evolutionary games where players could evolve other strategies, we see that costly punishment is not such a demerit to agents who assume this strategy. Results in this study show that in such games where agents are able to evolve other strategies based on payoffs, the average payoff of such networks are usually higher than networks without costly punishers. However, this is only possible in heterogeneous non-regular networks and all regular structures generally, homogenous or heterogeneous. In homogenous non-regular networks, the introduction of costly punishers brings down average payoff as shown in the results and also supported by the Dreber et al. experiment (15). The potential of costly punishers to evolve more profiting and sustaining strategies also accounts for the reason why this strategy is not observed to emerge as a sole evolutionary strategy. Hence, costly punishers are able to benefit and thrive in a network if they are able to decrease defection, consequently increasing the network's average payoffs, and then evolve to a sustaining strategy. By so doing, they indirectly increase their fitness and allow for favorable and sustainable strategies like cooperation to thrive. Thus, winners may actually punish – but not for too long.

There are studies that have shown the benefit of costly punishment in different networks (25-28); however, understanding why there is this discrepancy in the literature on the role of punishment, explaining why and when punishment becomes evolutionary beneficial is an important and needed insight. Figure 4 shows the rise in average payoff of a heterogeneous network of N = 25 agents. We can compare the average payoff of Figure 4 with that of Figure 6a. Figure 5a shows a homogenous regular structure, while Figure 5b shows a heterogeneous regular structure. We can compare the average payoff with those of Figure 3a and Figure 7a.

```
('Cope:', 9, ' Defe:', 91)
('Average payoff is', 8.07)
('Total agents', 25)
```

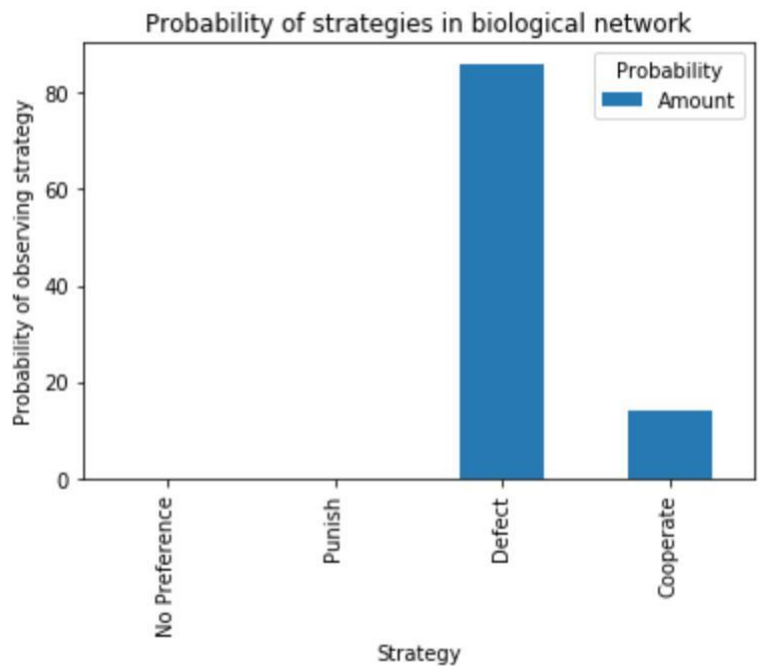

```
Counter({'Defect': 86, 'Cooperate': 14, 'No Preference': 0, 'Punish': 0})
```

**Figure 4:** The figure is the result of a heterogeneous network of N = 25 agents with costly punishers. The average payoff is 8.07 units which is 1.2 units higher than the heterogeneous network without costly punishers in Figure 6a.

```
('Cope:', 0, ' Defe:', 100)
('Average payoff is', 101.66)
('Total agents', 25)
```

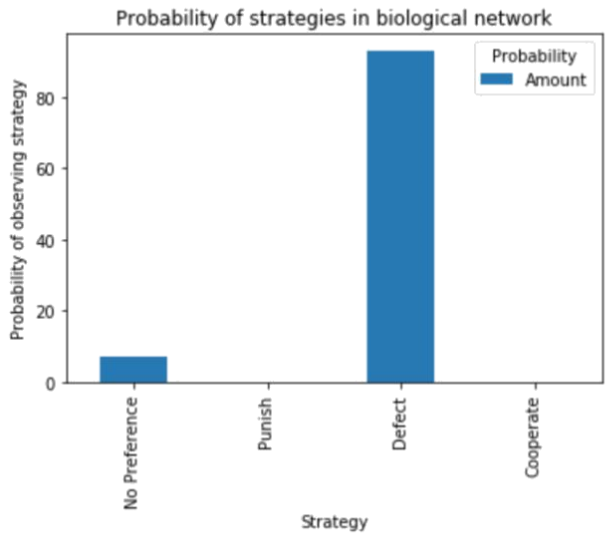

```
Counter({'Defect': 93, 'No Preference': 7, 'Punish': 0, 'Cooperate': 0})
```

**Figure 5a:** The figure gives the result of a homogeneous spatial lattice of N = 25. The average payoff is 101.66 units which is almost twice as large as the 51.74 units in Figure 3a.

```
('Cope:', 5, ' Defe:', 95)
('Average payoff is', 7.51)
('Total agents', 25)
```

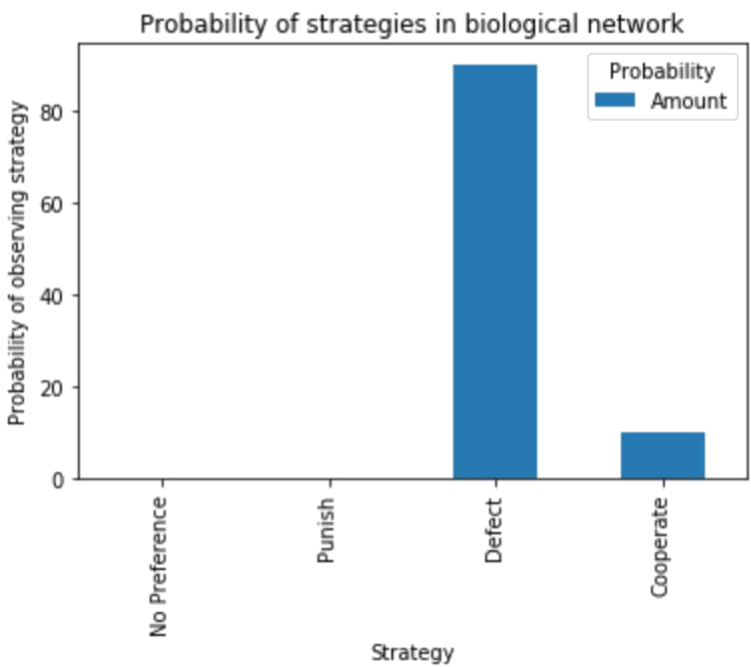

```
Counter({'Defect': 90, 'Cooperate': 10, 'No Preference': 0, 'Punish': 0})
```

**Figure 5b:** The figure shows the result of a heterogeneous regular spatial lattice of N = 25. The average payoff is 7.51 units which is about 2 units higher than that of the heterogeneous regular spatial lattice of N = 25 without costly punishers in Figure 7a that has an average payoff of 5.51 units. With higher N, this gap becomes more apparent.

### 3. Regular structures strip away non-sustainable strategies

In regular structures, punishers do not exist in the equilibrium state. This further emphasizes the anti-defective feature of a regular structure. The equilibrium state of regular networks for any particular N agents consist of only defectors and cooperators; however, non-regular networks may consist of punishers insofar there are defectors as well in that equilibrium state. A punisher would not emerge in an equilibrium state without defectors. There seems to be a coevolutionary relationship existing between punishers and defectors. This makes sense and seems to agree with numerous studies which have shown that costly punishment evolved as an evolutionary strategy against defection (5-6, 8, 25-28). Hence, it only follows that if there should be punishers in a stable network, then there should also be defectors. It has been discussed earlier how regular structures display anti-defective attributes but counteract the full effect of costly punishment. This could be due to the fact that costly punishment is most importantly, a non-sustaining strategy which further explains why it can never arise as a sole evolutionary strategy. The hypothesis is that regular structures impede non-sustainable strategies, which would explain their anti-defective feature and counteraction to the full anti-defective effect of costly punishment. This is even more pronounced in non-sustaining strategies because we observe that costly punishers cannot be found in the equilibrium state of regular networks.

## 4. Diversity in neighbors for regular and non-regular networks

In a previous study, we see that diversity in neighbors favors the emergence of cooperation, converting the equilibrium state to a cooperative strategy (5). It is also discussed in this study the need to tell apart 'the favoring of cooperation' and 'the disfavoring of defection' as these might mean the same thing in the local scale but is certainly not the same in the global scale. In the global scale we are looking at probabilities. A decrease in the probability of defection would not necessarily mean an increase in the probability of cooperation. As we have seen, strategies could also exist in equilibrium and this equilibrium state could be favored instead. In this study, we demonstrate that not only does diversity in neighbors favor the emergence of cooperation in homogenous non-regular networks with no costly punishers; it also disfavors defection, as supported by many other studies (29-34). In these experiments, we see how introducing diversity in neighbors in homogenous non-regular networks with no costly punishers and regular networks with costly punishers would lead to a decrease in defection. However, we see a rise in defection when diversity in neighbors is introduced in non-regular networks with costly punishers, as expected (5), and we also see a rise when diversity in neighbors is introduced in spatial lattices without costly punishers. This again illuminates the anti-defective and anti-punishing influence of regular structures like spatial lattices on non-sustainable strategies and as supported by research (35-37). A spatial lattice with costly punishers would produce the same results as a non-regular network without costly punishers because of this counteracting effect, and a spatial lattice without costly punishers, given it is already anti-defective, would produce the same results as a non-regular network with costly punishers in terms of the probability of defection rising as the sole evolutionary strategy. In all of these cases, on introduction of diversity in neighbors, of course, we do not observe an equilibrium state anymore. The equilibrium state, as well as the reduced fraction of defection for cases in which defection is disfavored, are all converted to cooperation. Many studies have presented conflicting opinions on the effect of diversity in neighbors due to this important oversight (29-34, 38-39). As earlier emphasized, biological networks are complex systems, and our observation of the influence of diversity in neighbors could also differ based on the type of network we are modeling. If the network is a structured one without costly punishers, then we would interpret diversity in neighbors as a parameter in favor of defection; whereas if we were looking at non-regular network without costly punishers, we would regard diversity in neighbors as a parameter that is anti-defective. Hence, merely by rearranging a network from non-regular to a more structured one, or vice versa, we can influence the effect of parameters like diversity. Figure 6 and 7 show results of simulation run when diversity in neighbors is introduced to networks N = 25 and N = 64. Figure 6 shows results for non-regular heterogeneous network while Figure 7 shows results for their respective regularized heterogeneous networks.

```
('Cope:', 0, ' Defe:', 100)
('Average payoff is', 6.87)
('Total agents', 25)
```

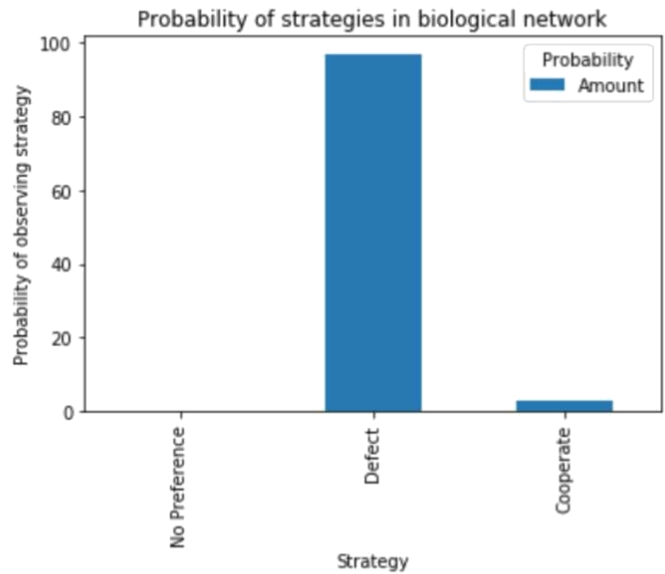

```
Counter({'Defect': 97, 'Cooperate': 3, 'No Preference': 0})
```

**Figure 6a:** In Figure 6a, we can clearly see a reduction in the probability of defection from 99% to 97% for N = 25, due to the influence of diversity in neighbors. There is also a subsequent emergence of cooperation as the entire equilibrium state is converted to the cooperative strategy.

```
('Cope:', 2, ' Defe:', 98)
('Average payoff is', 15.72)
('Total agents', 64)
```

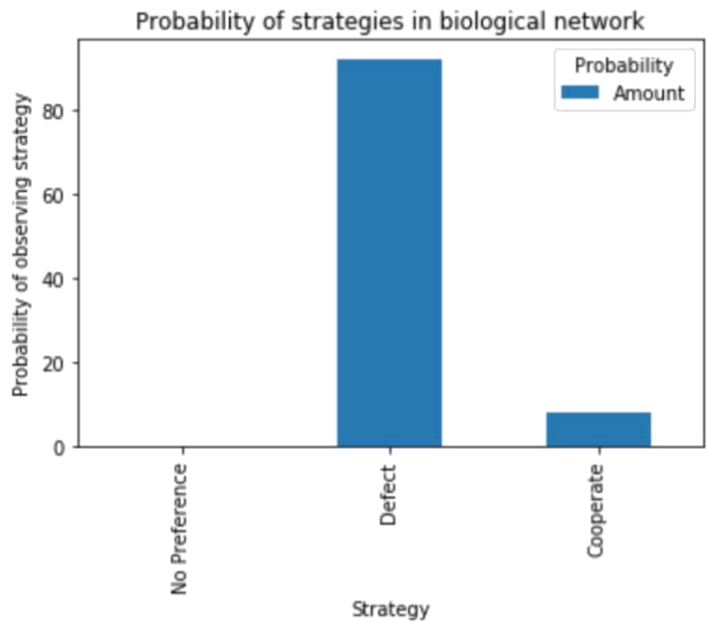

```
Counter({'Defect': 92, 'Cooperate': 8, 'No Preference': 0})
```

**Figure 6b:** We notice the same decrease in defection from 95% to 92% and a subsequent emergence in cooperation for N = 64.

```
('Cope:', 1, ' Defe:', 99)
('Average payoff is', 5.51)
('Total agents', 25)
```

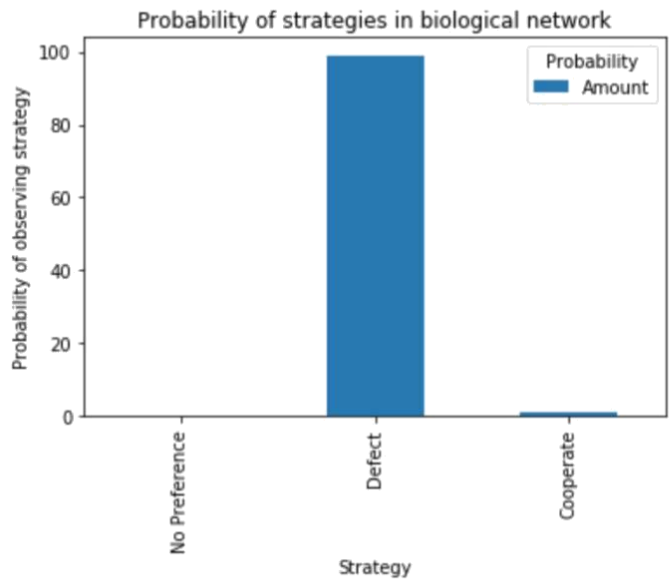

```
Counter({'Defect': 99, 'Cooperate': 1, 'No Preference': 0})
```

**Figure 7a:** We see a subsequent rise of defection from 97% in Figure 3 back to 99% when diversity in neighbors is introduced in a regular network for N = 25. This is a characteristic observable in networks with costly punishers, where the introduction of diversity in neighbors favors both defection and the emergence of cooperation. The homogenous non-regular network seems to have the same probabilities of defection with the heterogeneous regular network, and the homogenous regular network seems to have the same probabilities of defection with the heterogeneous non-regular network for N = 25.

```
('Cope:', 1, ' Defe:', 99)
('Average payoff is', 19.58)
('Total agents', 64)
```

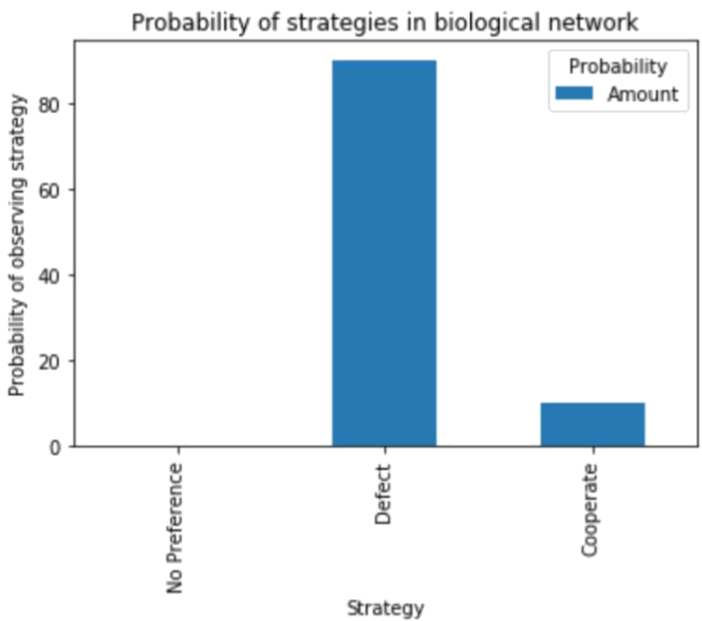

```
Counter({'Defect': 90, 'Cooperate': 10, 'No Preference': 0})
```

**Figure 7b:** The figure shows the same rise in the defective strategy for N = 64 from 85% in Figure 3 to 90%, although there is an emergence of cooperation.

## 5. Punishment on average payoff

Given that punishment is not unreasonably costly, this model demonstrates that it could actually yield evolutionary benefits measured in average payoff. The question on the purpose of punishment and why it evolved as a strategy in behavioral network interactions had for long puzzled the scientific community because although being sometimes altruistic and inhibiting defective behaviors, it was actually considered a strategy that reduces fitness and average payoff as mentioned earlier (15, 40-45). The phrase "winners don't punish" had been at the forefront of this belief. However, this model shows that this belief is only true for homogeneous non-regular networks. The case is however different, even the opposite, for heterogeneous non-regular networks and all regular networks. The model shows a significant decrease in payoffs for homogeneous non-regular networks with punishers as opposed to homogeneous non-regular networks without punishers. However, in the presence of network diversity in neighbors (heterogeneity), there is an immense boost in network average payoff in non-regular networks with punishers as opposed to both heterogeneous and homogeneous non-regular networks without punishers. As for regular networks, this model shows that regardless of the presence of diversity in neighbors, regular networks with punishers have significantly higher payoffs than regular networks without punishers (see Fig. 8). Tables I, II, II and IV are model results that show these relations as well as the probability of cooperation and defection for values of $N$ where,

$$N = n^2, \ for \ range \ n\colon 8 \le n \ \le 16 \ .$$

$N$  is a square here because we wish to accurately compare results of regular spatial lattices against non-regular networks. Given that average payoff and cooperation are influenced by $N$, which also defines the density of connectivity, $D_c$, these values must be kept constant the same for non-regular networks as well in order to compare them against the square lattices which are naturally squares.

| Regular Square Lattice With Punishers Agents (N) | Homogeneous | | | Heterogeneous | | |
|---|---|---|---|---|---|---|
| | D (%) | NP (%) | $P_{avg}$ | D (%) | C (%) | $P_{avg}$ |
| 64 | 79 | 21 | 146.51 | 65 | 35 | 45.66 |
| 81 | 67 | 33 | 184.44 | 58 | 42 | 53.63 |
| 100 | 71 | 29 | 128.07 | 63 | 37 | 45.35 |
| 121 | 47 | 53 | 248.45 | 63 | 37 | 89.01 |
| 144 | 49 | 51 | 182.31 | 23 | 77 | 121.52 |
| 169 | 46 | 54 | 184.05 | 27 | 73 | 113.78 |
| 196 | 41 | 59 | 161.72 | 18 | 82 | 135.94 |
| 225 | 34 | 66 | 217.72 | 14 | 86 | 149.12 |
| 256 | 36 | 64 | 182.17 | 9 | 91 | 161.08 |

**Table I:** Table I compares the network's average payoff, $P_{avg}$ , of a homogeneous regular square lattice with costly punishers (without diversity in neighbors) and a heterogeneous regular square lattice with costly punishers (with diversity in neighbors). "D" refers to the probability in percentage of defection arising as the sole evolutionary strategy; "NP" is the probability in percentage of the No Preference or equilibrium state arising as the sole evolutionary strategy; C is the probability in percentage of defection arising as the sole evolutionary strategy. The left side consists of only D and NP, while diversity is missing on the right side; this is because all NP's are converted to C's because of the influence of diversity in neighbors. The probabilities on the left side add up to 100%, and those on the right side as well add up to 100%, showing that these are the only evolved strategies after the game.

| Regular Square Lattice Without Punishers Agents (N) | Homogeneous | | | Heterogeneous | | |
| --- | --- | --- | --- | --- | --- | --- |
| | D (%) | NP (%) | $P_{avg}$ | D (%) | C (%) | $P_{avg}$ |
| 64 | 93 | 7 | 47.76 | 95 | 5 | 9.67 |
| 81 | 90 | 10 | 55.33 | 96 | 4 | 11.18 |
| 100 | 82 | 18 | 80.78 | 85 | 15 | 25.21 |
| 121 | 80 | 20 | 83.25 | 88 | 12 | 23.17 |
| 144 | 82 | 18 | 63.37 | 85 | 15 | 29.56 |
| 169 | 74 | 26 | 72.52 | 81 | 19 | 35.23 |
| 196 | 74 | 26 | 76.19 | 76 | 24 | 45.13 |
| 225 | 68 | 32 | 73.59 | 81 | 19 | 38.55 |
| 256 | 66 | 34 | 70.8 | 74 | 26 | 53.56 |

**Table II:** Table II compares the network's average payoff, $P_{avg}$, of a homogeneous regular square lattice without punishers (without diversity in neighbors) and a heterogeneous regular square lattice without punishers (with diversity in neighbors). Probabilities of defection emerging (D), cooperation emerging (C), or the equilibrium state emerging (NP) can be compared against Table I to clearly see the influence of costly punishment and diversity in neighbors on these strategies. Average payoffs, $P_{avg}$, are likewise influenced by these network parameters.

| Non-regular Network With Punishers Agents (N) | Homogeneous | | | Heterogeneous | | |
| --- | --- | --- | --- | --- | --- | --- |
| | D (%) | NP (%) | $P_{avg}$ | D (%) | C (%) | $P_{avg}$ |
| 64 | 22 | 78 | 8.97 | 60 | 40 | 48.91 |
| 81 | 10 | 90 | 20.11 | 52 | 48 | 67.15 |
| 100 | 8 | 92 | -51.08 | 47 | 53 | 85.61 |
| 121 | 2 | 98 | -8.39 | 33 | 67 | 109.2 |
| 144 | 3 | 97 | -42.47 | 30 | 70 | 130.94 |
| 169 | 0 | 100 | -14.44 | 29 | 71 | 127.92 |
| 196 | 0 | 100 | 5.99 | 23 | 77 | 143.17 |
| 225 | 0 | 100 | -20.09 | 19 | 81 | 167.25 |
| 256 | 0 | 100 | 5.67 | 16 | 84 | 171.31 |

**Table III:** Table III compares the network's average payoff, $P_{avg}$, of a homogeneous non-regular network with costly punishers (without diversity in neighbors) and a heterogeneous regular square lattice with costly punishers (with diversity in neighbors). Probabilities of defection emerging (D), cooperation emerging (C), or the equilibrium state emerging (NP) can be compared against Table I to clearly see the influence of network regularity on these strategies.

Average payoffs, $P_{avg}$, are likewise influenced by the network structure. It is important to observe the incredibly low average payoffs recorded for the homogeneous network. Comparing this against the right side, we see a huge difference in payoffs; whereas, the heterogeneous network has immensely higher payoffs, the homogeneous network has very low average payoffs. When compared against Table I, we see that for regular square lattices, the average payoffs are high regardless of the whether the network is homogeneous or heterogeneous.

| Non-regular Network Without Punishers Agents (N) | Homogeneous | | | Heterogeneous | | |
|---|---|---|---|---|---|---|
| | D (%) | NP (%) | $P_{avg}$ | D (%) | C (%) | $P_{avg}$ |
| 64 | 94 | 6 | 56.42 | 93 | 7 | 15.28 |
| 81 | 96 | 4 | 41.61 | 92 | 8 | 16.89 |
| 100 | 91 | 9 | 49.39 | 88 | 12 | 24.96 |
| 121 | 92 | 8 | 25.54 | 86 | 14 | 30.41 |
| 144 | 87 | 13 | 66.29 | 81 | 19 | 39.46 |
| 169 | 88 | 12 | 45.21 | 76 | 24 | 49.84 |
| 196 | 88 | 12 | 49.18 | 78 | 22 | 49.35 |
| 225 | 88 | 12 | 42.43 | 73 | 27 | 60.91 |
| 256 | 84 | 16 | 36.1 | 67 | 33 | 71.95 |

**Table IV:** Table IV compares the network's average payoff, $P_{avg}$, of a homogeneous non-regular network without punishers (without diversity in neighbors) and a heterogeneous regular square lattice without punishers (with diversity in neighbors). Probabilities of defection emerging (D), cooperation emerging (C), or the equilibrium state emerging (NP) can be compared against Table II to clearly see the influence of network regularity on these strategies. Average payoffs, $P_{avg}$, are likewise influenced by the network structure. Also comparing Table IV against Table III further illuminates the influence of costly punishment in non-regular networks, both in the presence and absence of diversity in neighbors. It is also very interesting to notice how introducing heterogeneity allows the average payoff to grow with increasing N for all tables.

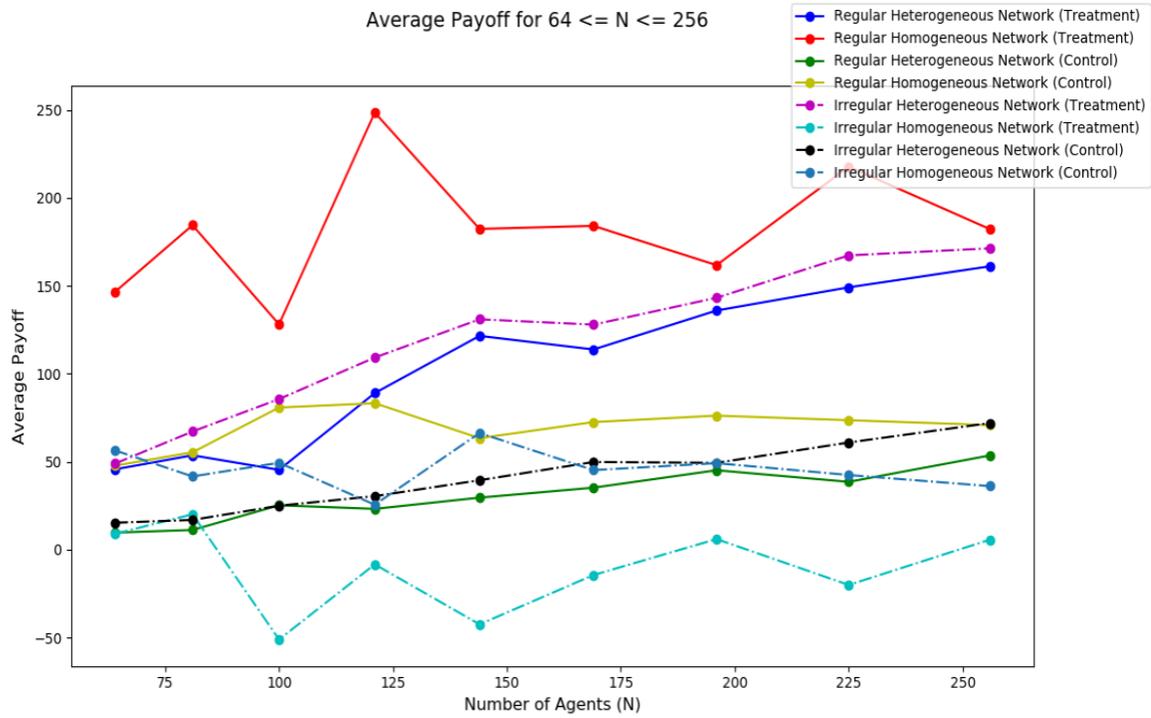

**Figure 8:** The figure depicts the difference in average payoff with homogeneous and heterogeneous regular and non-regular networks with increasing N; punishment is applied as treatment. This figure is a graphical representation of the data in Table IV.

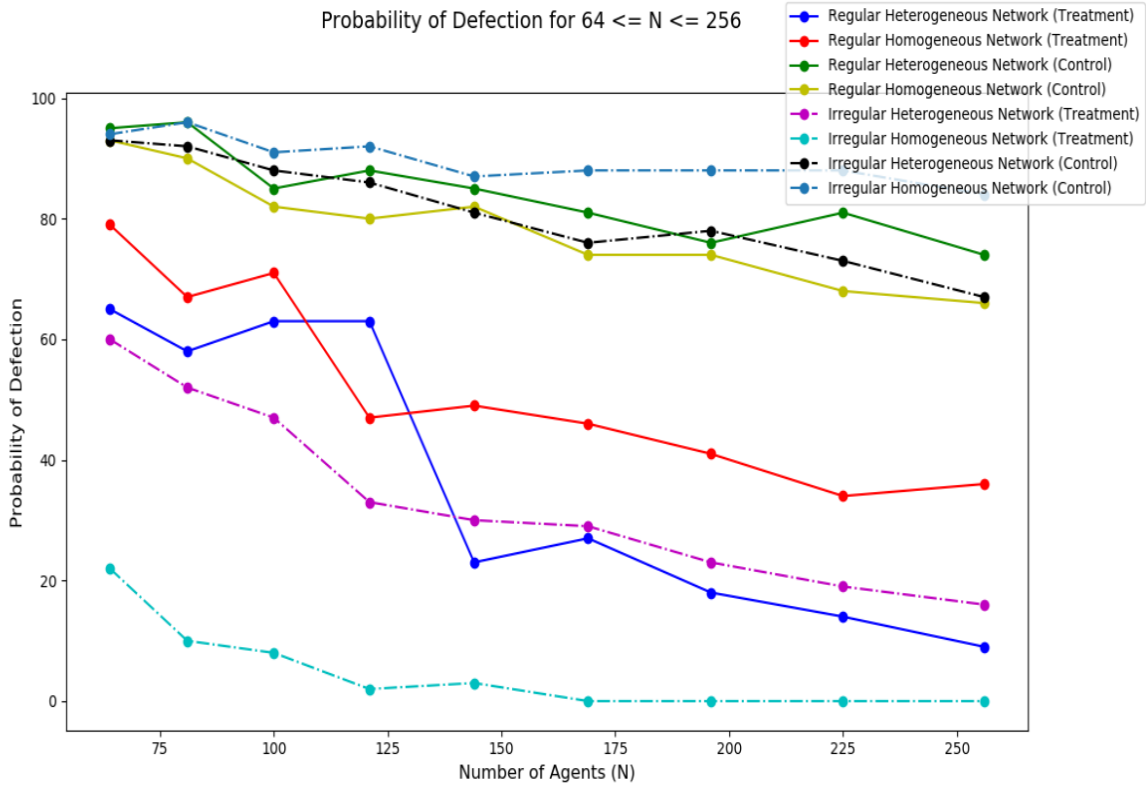

**Figure 9:** This figure shows the differences in defection for regular and non-regular homogeneous and heterogeneous networks in the presence (treatment) and absence (control) of costly punishers. We see a negative slope for each type of network because of the effect of an increasing $N$ on the density of connectivity with a constant $f$. We expect to observe a positive slope if the multiplying factor, $f$, were rather increasing on a constant $N$. We also see high probabilities of defection for networks without punishers as opposed to networks with punishers.

**Discussion**

We discover that myriads of changes in a biological network can arise from the rearrangement of non-regular network to a regular structure even though the degree of connectivity and the number of agents in the network remain constant. Essentially, moving around the edges to give a more structured network (where all agents have the same number of neighbors) is enough to disfavor defection, influence diversity in neighbors in puzzling ways, and counterpoise costly punishment. This may bring evolutionary explanations to why we find structure in many communities of living things – if communities (like colonies, filaments, and perhaps human societies) became more compact and structured to disfavor defection and allow for more cooperation, then evolutionary game theory can help answer the evolution of costly punishment in human societies (40-45).

## Why do people punish?

It has plagued the scientific community on why punishment has been preserved by natural selection (40-45). If according to some researchers, notably Dreber et al. (15), individuals with high payoffs do not punish and according to Table III we can see from the first panel that costly punishment reduces the network's average payoff, then why is such less fit strategy prominent in communities like human societies? Punishment is not only present in human societies, it also exist in other organisms (46-48). The results and experimental approach of this study suggests an answer to this question. As explained in Section 3, punishment never emerges at the global scale as an evolutionary strategy – especially in regular networks which act against non-sustainable strategies like punishment and defection. Section 2 and 5 however gives evidence to some hidden merits of costly punishment. Although the touchstone and predictor for natural selection in networks are the payoffs, many scientists have accepted that the answer may be found in the anti-defective effect of costly punishment in some network settings. This study shows us that aside from the anti-defective benefits as supported by Section 1, punishment can actually altruistically boost average payoff in all heterogeneous network and homogeneous regular networks. Now, the interesting question to consider is whether these populations are heterogeneous or regular – or both. Essentially, this means we would be able to answer the question on why costly punishment, as seen sometimes to be anti-progressive, is favored by natural selection in some populations if we know that these communities are regular or heterogeneous.

## Method Summary

The model used in this study is an agent-based model designed with Python. The algorithm follows the "imitate-the-best" strategy, where players imitate the strategy of the neighbor with the highest payoff given that the payoff of the imitated strategy is greater than theirs. Four agent-based models were designed to fully understand the relationship between regular structured networks and non-regular networks. The first two (numbered 1 and 2) are non-regular networks while the last two (numbered 3 and 4) are k-regular graphs. The Python code for these models is available here: https://github.com/ivanezeigbo/Regular-and-Non-regular-Biological-Networks. Payoffs which are calculated following Dreber et al (15) are updated synchronously. About equal number of cooperators, defectors, and punishers (for networks with costly punishment) initiate every game and each node is initially assigned a strategy randomly. When the number of agents is odd, the extra agent is always a defector (in networks without costly punishers) or a punisher. We have experimentally seen that the results are not significantly altered, if affected at all, in these cases. The model is designed to allow a user to visually follow the evolutionary progression in the networks thanks to the plot functions which display the state of the network diagrammatically with the help of the Networkx Python package.

# References

1. Boyd, R., & Richerson, P. (1992). Punishment allows the evolution of cooperation (or anything else) in sizable groups. *Ethology And Sociobiology*, *13*:171-195.
2. dos Santos, M., & Wedekind, C. (2015). Reputation based on punishment rather than generosity allows for evolution of cooperation in sizable groups. *Evolution And Human Behavior*, *36*:59-64.
3. dos Santos, M., Rankin, D., & Wedekind, C. (2013). Human Cooperation Based on Punishment Reputation. *Evolution*, *67*:2446-2450.
4. Hauert, C., Traulsen, A., Brandt, H., Nowak, M., & Sigmund, K. (2007). Via Freedom to Coercion: The Emergence of Costly Punishment. *Science*, *316*:1905-1907.
5. Ezeigbo, I. (2017). How costly punishment, diversity, and density of connectivity influence cooperation in a biological network. *Scientific Reports*, *7*(1).
6. Boyd, R., Gintis, H. and Bowles, S. Coordinated Punishment of Defectors Sustains Cooperation and Can Proliferate When Rare. (2010) *Science*, 328:617-620.
7. Ohtsuki, H., Hauert, C., Lieberman, E., & Nowak, M. (2006). A simple rule for the evolution of cooperation on graphs and social networks. *Nature*, *441*:502-505.
8. Krasnow, M., Cosmides, L., Pedersen, E. and Tooby, J. (2012) What Are Punishment and Reputation for? *PLoS ONE*, 7:e45662.
9. Vukov, J., Pinheiro, F., Santos, F. and Pacheco, J. (2013) Reward from Punishment Does Not Emerge at All Costs. *PLoS Computational Biology*, 9:e1002868.
10. Dreber, A., Fudenberg, D., & Rand, D. (2011). Who Cooperates in Repeated Games? *SSRN Electronic Journal*.
11. Nadell, C., Foster, K. and Xavier, J. (2010) Emergence of Spatial Structure in Cell Groups and the Evolution of Cooperation. *PLoS Computational Biology*, 6:e1000716.
12. Roca, C., Cuesta, J. and Sánchez, A. (2009) Effect of spatial structure on the evolution of cooperation. *Physical Review E*, 80(4).
13. Rand, D., & Nowak, M. (2013). Human cooperation. *Trends In Cognitive Sciences*, *17*:413-425.
14. Verbruggen, E., El Mouden, C., Jansa, J., Akkermans, G., Bücking, H., West, S., & Kiers, E. (2012). Spatial Structure and Interspecific Cooperation: Theory and an Empirical Test Using the Mycorrhizal Mutualism. *The American Naturalist*, *179*:E133-E146.
15. Dreber, A., Rand, D., Fudenberg, D., & Nowak, M. (2008). Winners don't punish. *Nature*, *452*:348-351.
16. Zhu, C., Sun, S., Wang, L., Ding, S., Wang, J., & Xia, C. (2014). Promotion of cooperation due to diversity of players in the spatial public goods game with increasing neighborhood size. *Physica A: Statistical Mechanics And Its Applications*, *406*:145-154.
17. Shi, D., Zhuang, Y., & Wang, B. (2010). Group diversity promotes cooperation in the spatial public goods game. *EPL (Europhysics Letters)*, *90*:58003.
18. Santos, F., Santos, M., & Pacheco, J. (2008). Social diversity promotes the emergence of cooperation in public goods games. *Nature*, *454*: 213-216.
19. Pfattheicher, S., Böhm, R., & Kesberg, R. (2018). The Advantage of Democratic Peer Punishment in Sustaining Cooperation within Groups. *Journal Of Behavioral Decision Making*.

20. Rand, D., Ohtsuki, H., & Nowak, M. (2009). Direct reciprocity with costly punishment: Generous tit-for-tat prevails. *Journal Of Theoretical Biology*, *256*:45-57.

21. Santos, M., Rankin, D., & Wedekind, C. (2010). The evolution of punishment through reputation. *Proceedings Of The Royal Society B: Biological Sciences*, *278*:371-377.

22. Suzuki, S., & Akiyama, E. (2005). Reputation and the evolution of cooperation in sizable groups. *Proceedings Of The Royal Society B: Biological Sciences*, *272*:1373-1377.

23. Hauert, C. and Doebeli, M. (2004) Spatial structure often inhibits the evolution of cooperation in the snowdrift game. *Nature Materials*, 428:643-646.

24. Wu, J., Zhang, B., Zhou, Z., He, Q., Zheng, X., Cressman, R., & Tao, Y. (2009). Costly punishment does not always increase cooperation. *Proceedings Of The National Academy Of Sciences*, *106*:17448-17451.

25. Deng, K., Li, Z., Kurokawa, S. and Chu, T. (2012) Rare but severe concerted punishment that favors cooperation. *Theoretical Population Biology*, 81:284-291.

26. Boyd, R., Gintis, H., Bowles, S. and Richerson, P. (2003) The evolution of altruistic punishment. *Proceedings of the National Academy of Sciences*, 100:3531-3535.

27. Perc, M., Jordan, J., Rand, D., Wang, Z., Boccaletti, S., & Szolnoki, A. (2017) Statistical physics of human cooperation. *Physics Reports*, *687*:1-51.

28. Helbing, D., Szolnoki, A., Perc, M., & Szabó, G. (2010). Punish, but not too hard: how costly punishment spreads in the spatial public goods game. *New Journal Of Physics*, *12*:083005

29. Perc, M. Does strong heterogeneity promote cooperation by group interactions? (2011) *New Journal Of Physics*, *13*:123027.

30. Santos, F., Pinheiro, F., Lenaerts, T., & Pacheco, J. (2012) The role of diversity in the evolution of cooperation. *Journal Of Theoretical Biology*, *299*:88-96.

31. Szolnoki, A., Perc, M. and Szabó, G. (2008) Diversity of reproduction rate supports cooperation in the prisoner's dilemma game on complex networks. *The European Physical Journal B*, 61:505-509.

32. Perc, M., & Szolnoki, A. (2008). Social diversity and promotion of cooperation in the spatial prisoner's dilemma game. *Physical Review E*, *77*(1).

33. Yang, H., Wang, W., Wu, Z., Lai, Y., & Wang, B. (2009). Diversity-optimized cooperation on complex networks. *Physical Review E*, *79*(5).

34. Zhu, P., & Wei, G. (2014). Stochastic Heterogeneous Interaction Promotes Cooperation in Spatial Prisoner's Dilemma Game. *Plos ONE*, *9*:e95169.

35. Perc, M., & Szolnoki, A. (2012) Self-organization of punishment in structured populations. *New Journal Of Physics*, *14*:043013.

36. Grujić, J., Fosco, C., Araujo, L., Cuesta, J. and Sánchez, A. (2010) Social Experiments in the Mesoscale: Humans Playing a Spatial Prisoner's Dilemma. *PLoS ONE*, 5:e13749.

37. Nowak, M. and May, R. (1992) Evolutionary games and spatial chaos. *Nature*, 359:826-829.

38. Melamed, D. and Simpson, B. (2016) Strong ties promote the evolution of cooperation in dynamic networks. *Social Networks*, 45:32-44.

39. Gracia-Lazaro, C., Ferrer, A., Ruiz, G., Tarancon, A., Cuesta, J., Sanchez, A. and Moreno, Y. (2012) Heterogeneous networks do not promote cooperation when humans play a Prisoner's Dilemma. *Proceedings of the National Academy of Sciences*, 109(32):12922-12926.

40. Fehr, E., & Fischbacher, U. (2004). Social norms and human cooperation. *Trends In Cognitive Sciences*, *8*:185-190.

41. Fehr, E., Fischbacher, U., & Gächter, S. (2002). Strong reciprocity, human cooperation, and the enforcement of social norms. *Human Nature*, *13*:1-25.

42. Gardner, A., & West, S. (2004). Cooperation and Punishment, Especially in Humans. *The American Naturalist*, *164*:753-764.

43. Rankin, D., dos Santos, M., & Wedekind, C. (2009). The evolutionary significance of costly punishment is still to be demonstrated. *Proceedings Of The National Academy Of Sciences*, *106*:E135-E135.

44. Henrich, J., McElreath, R., Barr, A., Ensminger, J. and Bolyanatz, A. (2006) Costly Punishment Across Human Societies. *Science*, 312(5781): 1767-1770.

45. Sigmund, K. (2007) Punish or perish? Retaliation and collaboration among humans. *Trends In Ecology & Evolution*, *22*: 593-600.

46. Fu, F., & Wang, L. (2008) Coevolutionary dynamics of opinions and networks: From diversity to uniformity. *Physical Review E*, *78*(1).

47. Nowak, M. Five Rules for the Evolution of Cooperation. (2006) *Science*, 314(5805):1560-1563.

48. Doebeli, M. and Knowlton, N. (1998) The evolution of interspecific mutualisms. *Proceedings of the National Academy of Sciences*, 95(15):8676-8680.

49. Killingback, T., Doebeli, M. and Knowlton, N. (1999) Variable investment, the Continuous Prisoner's Dilemma, and the origin of cooperation. *Proceedings of the Royal Society B: Biological Sciences*, 266:1723-1728.